\documentclass[twocolumn,prb,showkeys,showpacs,preprintnumbers,amsmath,amssymb]{revtex4}
\usepackage{graphicx}% Include figure files
\usepackage{dcolumn}% Align table columns on decimal point
\usepackage{bm}% bold math

\begin{document} 

\title{Ab-initio prediction of a new multiferroic with
large polarization and magnetization}

\author{Pio Baettig}
\altaffiliation[Also at ]{Chemistry Department, Universit\'{e} de Fribourg, P\'{e}rolles, CH-1700 Fribourg, Switzerland.}
\author{Nicola Spaldin}%
\affiliation{%
Materials Department\\
University of California, Santa Barbara, California 93106-5050, USA\\
}%

\date{\today}
\newcommand{\BFCO}{Bi$_2$FeCrO$_6$}

  %%%%%%%%%%%%%%%%% END OF PREAMBLE %%%%%%%%%%%%%%%%
\hyphenation{%
%A
%B
%C
%D
%E
%F
%G
%H
%L
%M
%N
%O
%P
%Q
%R
rea-sons
%S
%T
%U
%V
%W
%X
%Y
%Z
}

\begin{abstract}
We describe the design of a new magnetic ferroelectric with large spontaneous magnetization 
and polarization using first-principles density functional theory. The usual difficulties 
associated with the production of robustly-insulating ferromagnets are circumvented by 
incorporating the magnetism through {\it ferri-}magnetic behavior. We show that
the the ordered perovskite \BFCO\ will have a polarization of $\sim$80 $\mu$C/cm$^2$,
a piezoelectric coefficient of 283 $\mu$C/cm$^{2}$, and a magnetization of $\sim$160 
emu/cm$^3$ (2 $\mu_B$ per formula unit), far exceeding the properties of any known multiferroic.
\end{abstract}

\maketitle

Multifunctional materials that combine a spontaneous magnetization with a 
ferroelectric polarization are of tremendous fundamental and practical interest
\cite{Fiebig_Nature:2002,Zheng/Ramesh_BTO-CFO:2004,Sudak:2004,Schmid}.
For example they have high dielectric permittivity and high magnetic
permeability, and could therefore replace the inductor and capacitor in resonant
circuits with a single component, further miniaturizing portable cellular technologies. 
Strong coupling between the polarization and magnetization would allow ferroelectric 
data storage combined with a magnetic read. And the ability to tune or switch 
the magnetic properties with an electric field and vice versa could lead to 
as-yet-unanticipated developments in conventional devices such as transducers.
However a single phase material with large and robust magnetization and 
polarization has not been previously identified; known ferromagnetic ferroelectrics 
tend to have low magnetic Curie temperatures\cite{Kimura_Nature:2003}, are often 
weak ferromagnets\cite{Fox_Scott_1977}, or are not strong enough insulators to
sustain a ferroelectric polarization at room temperature\cite{dosSantos2}. 

The coexistence of ferromagnetism and ferroelectricity is difficult to 
achieve for many reasons\cite{hill,Khomskii}. First, and most fundamentally,
the off-centering of transition metal ions which creates the electric 
polarization in conventional ferroelectrics is driven by a second-order 
Jahn-Teller distortion which requires a formally empty $d$-electron 
configuration\cite{hill}. In contrast, ferromagnetism requires unpaired 
electrons, which in many materials are provided by $d$ electrons on transition
metal ions. So the
coexistence of the two phenomena, although not prohibited by any
physical law or symmetry consideration, is discouraged by the local
chemistry that favors one or the other but not both. In practice, 
alternative mechanisms for introducing off-centering on ions other
than the magnetic transition metal ions, such as stereochemically-active
lone pairs\cite{Seshadri_Hill} or geometrically-driven distortions\cite{YMnO3_Nature},
can be used to circumvent this restriction.
Second, and important from a practical point of view,
is the fact that ferroelectrics must be robust insulators, so that an applied
field can reorient the spontaneous polarization without causing conductivity. 
In practice, many ferromagnetic materials are
metallic, and magnetic insulators tend to show {\it anti}ferromagnetic
coupling between the magnetic ions. Finally, magnetic transition metal
ions tend to be fairly easy to oxidize or reduce, and are associated with 
multiple valence states accompanied by anion non-stoichiometry.
 So those few ferromagnetic insulators that do exist
in fact show hopping conductivity at room temperature\cite{dosSantos2}.

In this work we propose a new class of robustly-insulating magnetic ferroelectrics
in which all three of these limitations are circumvented. We use the $6s^2$ lone pair
on Bi$^{3+}$ ions, well-established to be the source of the ferroelectricity in
the multiferroics BiMnO$_3$\cite{Seshadri_Hill} and BiFeO$_3$\cite{Neaton_etc:unpublished}, to 
introduce the off-center distortion.
To avoid the difficulty of finding insulating ferromagnetic behavior, we instead
choose to introduce spontaneous magnetization via {\it ferrimagnetism.} In 
ferrimagnets, the coupling between neighboring magnetic moments is robustly 
antiferromagnetic, but two antiferromagnetic sublattices of different magnetizations 
are incorporated to provide a net
magnetization. Because the antiferromagnetic superexchange is so strong, the
magnetic Curie temperatures in ferrimagnets are usually far above room temperature.
In addition, to promote strongly insulating behavior, we choose transition 
metal ions which are resistant to oxidation and reduction: 
$d^3$ Cr$^{3+}$, in which the up-spin t$_{2g}$-shell is filled, and high-spin $d^5$
Fe$^{3+}$, with its completely filled up-spin manifold. 

\begin{figure}
\includegraphics*[width=1.0\columnwidth]{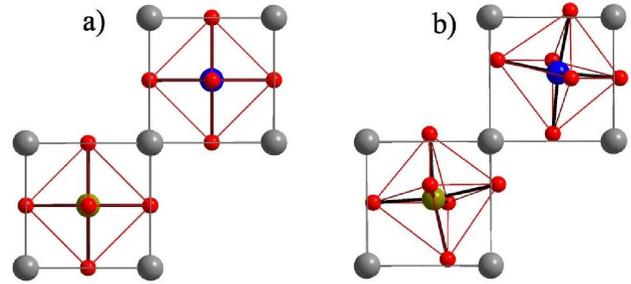}
\caption{Comparison of ideal cubic perovskite-structure \BFCO\ (a) with our calculated
ground state structure (b), looking down the $y$ axis of the ideal cubic unit cell. (111) 
planes of Fe ions (light gray) alternate with planes of Cr ions (dark gray) along the
[111] direction. Each transition metal is surrounded by an octahedron of oxygen anions
(black), and the Bi ions (medium gray) occupy the corners of the ideal cubic unit cells.
The transition from the cubic to the ground state structure consists of two distortions:
rigid rotations of adjacent oxygen octahedra in opposite directions around the [111]
axis, and displacements of the anion cages relative to the cations in the [111] direction. }
\label{structure}
\end{figure}

Having selected our cationic composition (Bi$^{3+}$, Cr$^{3+}$, Fe$^{3+}$), our 
next step is to choose a trial arrangement of the ions, and to calculate the 
crystal structure and magnetic properties. Motivated by the report of the possibility
of single-atomic-layer superlattice ordering in double perovskite 
La$_2$FeCrO$_6$\cite{la2fecro6}, we choose as our trial compound the ordered perovskite 
\BFCO, with planes of Fe and Cr ions alternating in the [111] direction; this corresponds
to a rocksalt ordering of the Fe- and Cr-octahedra, as shown in Figure~\ref{structure} (a).
We use density functional theory within the LDA+$U$ method\cite{Anisimov:1997}, as 
implemented in the VASP package\cite{Kresse/Furthmueller_VASP1:1996}, to determine 
the properties. This extension to the usual local spin density approximation
(LSDA) is necessary, because the LSDA gives qualitatively incorrect results for
systems with a strong local correlation. In our case, LSDA predicts that \BFCO\ should
be a metal containing low-spin Fe ions, instead of the high-spin insulator obtained
within the LDA+$U$ method. Technical details include a $U$ value of 4 eV, a $J$
value of 0.8 eV, a 6$\times$6$\times$6 k-point grid for the two formula-unit rhombohedral 
unit cell, the projector-augmented plane wave (PAW) method with default
pseudopotentials\cite{Kresse/Joubert_PAW:1999} and an energy cut-off of 450 eV.
$U$=4 eV  is the smallest value that gives an insulating centrosymmetric structure; 
this represents a lower limit for $U$, which is not inconsistent with the experimental estimation
for LaCrO$_3$ and LaFeO$_3$\cite{Arima93}.

\begin{figure}
\includegraphics*[width=0.5\columnwidth]{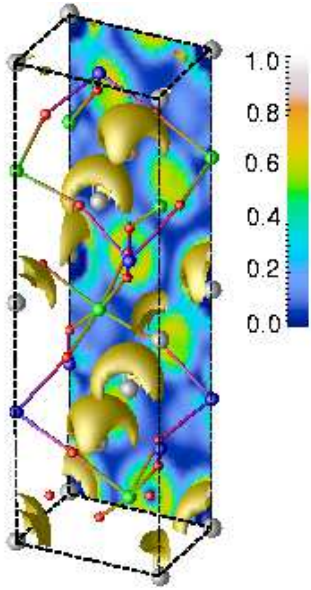}
\caption{Calculated electron localization function (ELF), at a value of 0.65, for \BFCO. The polar
[111] axis is oriented vertically (the Bi - Fe - Bi - Cr - Bi ordering can be seen clearly along
the front right edge of the cell). The lobes are regions of high electron localization 
associated with the lone pairs of electrons on the Bi ions. A contour plot of the electron
localization in a slice at the back of the cell is also shown. The ELF was generated 
using the STUTTGART TB-LMTO-ASA-code \cite{Andersen_LMTO:1975} with the LSDA formalism.}
\label{elf}
\end{figure}

First we calculate the lowest-energy crystal structure using the standard method 
of adjusting the positions of the atoms to minimize their Hellmann-Feynman forces.
To ensure that our calculations explore all likely symmetries in a reasonable 
computational time, we use a range of starting structures for the optimization, 
which we obtain by freezing in combinations of the unstable phonon modes of 
BiCrO$_3$\cite{Hill_Pio}. We find that the lowest energy structure 
has $R3$ symmetry (spacegroup 146) and exhibits alternating rotations of the
oxygen octahedra along 
the [111] direction, combined with relative displacements of the anions and cations 
along [111] (see Figure~\ref{structure}). This symmetry permits the existence of 
ferroelectricity, and indeed we calculate a ferroelectric polarization (using the 
Berry phase formalism\cite{kv,ksv,resta2}) of  79.6 $\mu$C/cm$^{2}$. The ferroelectric 
distortion is driven by the well-established stereochemical activity on the Bi lone 
pair\cite{Seshadri_Hill} as can be seen clearly in the calculated electron localization 
function\cite{silvi} (Figure~\ref{elf}). We calculate a piezoelectric coefficient along 
[111] of 283 $\mu$C/cm$^{2}$, comparable to that of PbTiO$_3$\cite{saghi} in which the 
piezoelectric coefficient along the polar $z$ axis, $e_{33}$, is 323 $\mu$C/cm$^2$.

Next we determine the magnetic ordering. We test all the likely possibilities that 
are accessible in our computations: ferromagnetic and antiferromagnetic ordering 
of the Fe planes and Cr planes, coupled both ferromagnetically and antiferromagnetically 
to each other. (Note that our super-cell technique does not allow us to study less likely
long-wavelength
magnetic orderings, and so we have not explored more obscure magnetic arrangements,
or non-collinear spin structures). Here we find that the most favorable magnetic
ordering consists of ferromagnetic planes of Fe and Cr, that are antiferromagnetically
coupled to each other. Since the magnetic moments of Fe and Cr are different (5 $\mu_B$
for Fe$^{3+}$ and 3 $\mu_B$ for Cr$^{3+}$) this arrangement results in a net magnetic
ordering of 2 $\mu_B$ per Fe-Cr pair, which corresponds to 160.53 emu/cm$^3$.

\begin{figure}
\includegraphics*[width=0.8\columnwidth]{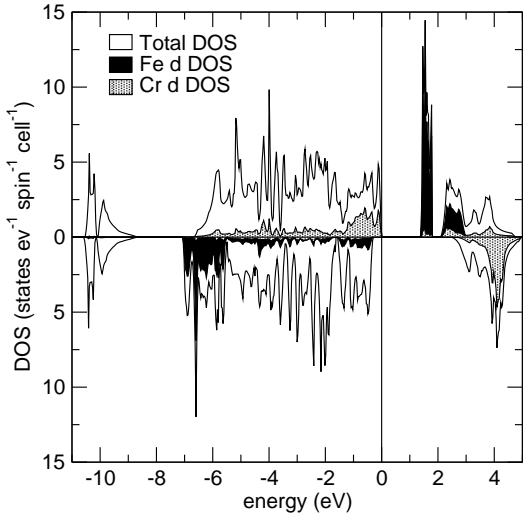}
  \caption{Calculated density of states (DOS) of \BFCO. The top of the valence band
is set to 0 eV. The black line shows the total density of states, and the dark and light shading
the contributions from the Fe $3d$ and Cr $3d$ states respectively. The states at $\sim$ 
-10 eV are the Bi $6s$ states, and the unshaded states in the
broad band below the Fermi energy have largely O $2p$ character.}
\label{dos}
  \end{figure}

Thus we predict that the properties of \BFCO\ will surpass all previous
reports for coexisting magnetism and ferroelectricity. The ferroelectric
polarization is comparable to the best known ferroelectrics (the value for 
PbTiO$_3$, for example, is $\sim$ 75 $\mu$C/cm$^2$ \cite{gavri}), the piezoelectric
response is substantial, and the large atomic
displacements and polarization reflect a high ferroelectric Curie temperature.
The material is insulating; we
have chosen transition metal ionic configurations that are the most robust to
oxidation and reduction. And the spontaneous magnetization is comparable to that 
of the cubic ferrites (for example the magnetization of nickel ferrite is 
$\sim$ 270 emu/cm$^{3}$).
Finally we mention that, although we deliberately designed
our system to be a robust insulator, if carriers are introduced either by doping
or by optical excitation, the transport and optical properties could be useful
for spintronic applications. The calculated density of states, shown in Figure~\ref{dos}, 
indicates that both the top of the valence band and the bottom of the conduction band 
are 100\% spin-polarized, with electrons confined to narrow, up-spin Fe $3d$ bands, 
and holes in broader, up-spin Cr $3d$ - O $2p$ hybrids.

Our work illustrates the viability of electronic structure methods in
the design of new and multifunctional materials and offers a synthetic 
challenge which we hope will stimulate further research on this emerging 
class of multiferroic materials. 

\paragraph*{Acknowledgments}
We thank Claude Ederer and Claude Daul for fruitful discussions regarding our
calculations.

\newpage

\bibliography{Nicola}

\newpage

  %%%%%%%%%%%%%%%%%%%%%%%%%%%%%%%%%%%%%%%%%%%%

  \end{document}